\documentclass[preprint,bm,cite]{epl}

\usepackage{graphicx}
\usepackage{amssymb,amsmath}

\shrinkacknowledgements{1} \shrinkfloats{2}

\newcommand{\smeq}{\! = \!}

\newcommand{\smpl}{\! + \!}
\newcommand{\smmi}{\! - \!}

\newcommand{\ve}{\varepsilon}

\newcommand{\Ef}{E_{\mathrm{F}}}

\newcommand{\be}{\begin{equation}}
\newcommand{\ee}{\end{equation}}
\newcommand{\bea}{\begin{eqnarray}}
\newcommand{\eea}{\end{eqnarray}}
\newcommand{\Ha}{{\hat H}}

\newcommand{\up}{\uparrow}
\newcommand{\dn}{\downarrow}

\newcommand{\dm}{\delta_\mathrm{m}}
\newcommand{\dM}{\delta_\mathrm{M}}
\newcommand{\dEf}{\delta{\Ef}}
\newcommand{\dk}{\delta k}

\title{Anisotropy in Ferromagnetic Nanoparticles: Level-to-Level Fluctuations of a Collective Effect}
\shorttitle{Anisotropy in Ferromagnetic Nanoparticles}
\author{Gonzalo Usaj\inst{1} \and Harold U. Baranger\inst{2}}
\shortauthor{G. Usaj \and H. U. Baranger}
\institute{
\inst{1} Instituto Balseiro and Centro At\'{o}mico Bariloche, (8400) Bariloche, Argentina\\
\inst{2} Department of Physics, Duke University, Durham North Carolina 27708-0305\\
}
\pacs{73.22.-f}{Electronic structure of nanoscale materials}
\pacs{73.23.Hk}{Coulomb blockade; single-electron tunneling}
\pacs{75.75.+a}{Magnetic properties of nanostructures}

\begin{document}
\maketitle

\begin{abstract}
We calculate the mesoscopic fluctuations of the magnetic anisotropy of ferromagnetic nanoparticles; that is, the effect of single-particle interference on the direction of the collective magnetic moment. A microscopic spin-orbit Hamiltonian considered as a perturbation of the much stronger exchange interaction first yields an explicit expression for the anisotropy tensor. Then, assuming a simple random matrix model for the spin-orbit coupling allows us to describe the fluctuation of such a tensor analytically. In the case of uniaxial anisotropy, we calculate the distribution of the anisotropy constant for a given number of electrons, and its variation upon increasing this number by one. The magnitude of the latter scales inversely with the number of atoms in the particle and is sufficient to account for the experimental data.
\end{abstract}

In a remarkable step toward a better understanding of ferromagnetism at the nanometer scale, individual electronic states within ferromagnetic nanoparticles were recently observed using electron tunneling spectroscopy \cite{GueronDMR99,DeshmukhKGBPDR01}.  Several features of the tunneling excitation spectra could not be explained within the simple electron-in-a-box picture, unlike the normal metal case \cite{vanDelftR01}.  Although this ``failure'' was anticipated due to the many-body character of ferromagnetism, several particular aspects were largely unexpected: (1) the non-linear magnetic field dependence of the energy levels close to the switching field ($ B_\mathrm{sw} $, the field at which the magnetization suddenly changes its orientation), (2) very small g-factors, and (3) excitation energies much smaller than the mean level spacing associated with the finite particle size.  While we briefly return to the latter two in closing, we focus on the first effect as representative of the peculiarities of nanoscale ferromagnetism.

A non-zero $ B_\mathrm{sw}$ implies the presence of a magnetic
anisotropy energy barrier: the (large) magnetic moment of the
particle points in a preferred direction. This preference
originates from spin-orbit coupling and/or the shape of the grain.
At $ B_\mathrm{sw}$, there is an abrupt change in the grain's
total energy, leading to ``jumps" in the tunneling spectrum.
Refs. \cite{GueronDMR99,DeshmukhKGBPDR01,KleffDDR01} show that
such jumps can be described by a simple uniaxial anisotropy model:
$ \Ha_\mathrm{A}= -k\, \hat{S}_z^2/S$, where $ k$ is the
anisotropy constant---which is an \textit{intensive} quantity in
the bulk---$ S $ is the total spin and $ \hat{z} $ is the
anisotropy axis.  However, to account for the non-linear behavior
and the magnitude of the jumps, it was crucial to \textit{assume}
that $ k $ depends on both the number of electrons in the
nanoparticle \textit{and} the particular many-body states involved
in each electronic transition. A change of $ 1$-$3\% $ upon adding
one electron in a nanoparticle with roughly $1000$ atoms was
enough to account for the experiments. Although fluctuation of
electronic properties is expected at the nanoscale, a microscopic
connection to anisotropy is needed.

In the last years, several works have aimed to understand the origin of the additional resonances \cite{DeshmukhKGBPDR01,KleffDDR01,CanaliM00,KleffD02,CehovinCM03} or the fluctuation of the anisotropy \cite{CehovinCM02,BrouwerG04}.  While the former remains controversial, the latter seems to be accounted for (in nanoparticles with up to $ 260 $ atoms) by the self-consistent numerical calculation done in Ref.~\cite{CehovinCM02}. One may wonder, however, if a simpler model, suitable for an analytic treatment of larger nanoparticles, can give similar results while providing additional physical insight.

In this work, we first derive an expression for the magnetic anisotropy tensor in terms of the matrix elements of the spin-orbit (SO) Hamiltonian using a simple Hamiltonian model for the ferromagnetic nanoparticle. Then, by using random matrix theory (RMT) to account for the fluctuations of the SO matrix elements, we calculate the probability distribution of the magnetic anisotropy constant for the case of uniaxial anisotropy. For nanoparticles with $ N_a\!\simeq\!1000 $ atoms, we find that fluctuations of $ 1\% $ are typical.
Furthermore, the fluctuations scale inversely with $N_a$ so that our results agree with those of Ref.~\cite{CehovinCM02} for smaller nanoparticles.

Ferromagnetism originates in the electron-electron interaction---in particular, the exchange interaction.
The simplest model  Hamiltonian, then, reads
\be
\Ha = \Ha_{\mathrm{S}} + \Ha_{\mathrm{Z}} +\Ha_{\mathrm{SO}}
     =\sum_{m,\sigma}{\ve^{}_{m}}\,\hat{n}^{}_{m\sigma}- J_{S}{\vec S}^2
            + \Ha_{\mathrm{Z}} +\Ha_{\mathrm{SO}}\,,
\label{Hamiltonian}
\ee
where $\{\ve_{m}\}$ are the single particle energies, ${\vec S}$ is the total spin operator, and $J_{S}$ is the exchange constant.
The latter produces the large spin of the nanoparticle (typically $S\approx1000$ \cite{GueronDMR99}); notice that this model assumes that the magnetic moment is coherent
and uniform \cite{CehovinCM02}.
The second and third terms in Eq. (\ref{Hamiltonian}) describe the
Zeeman energy, $H_{\mathrm{Z}}= -g\mu_\mathrm{B} \vec{B}\!\cdot\!\vec{S}$, and the SO coupling,
\be
\Ha_{\mathrm{SO}}= \mathrm{i}\sum_{n,m}\sum_{\sigma,\sigma'} (
   {\vec{A}}_{nm} \cdot {\vec \sigma}_{\sigma\sigma'} )
\,c^{\dagger}_{n\sigma}c^{}_{m\sigma'} \, ,
\ee
respectively. $ \Ha_{\mathrm{SO}} $ is responsible for the
magnetic anisotropy---we neglect the contribution of the shape
a\-ni\-so\-tro\-py (magnetostatic) as its relative contribution
decreases for smaller particles and so is very small at the
nanoscale \cite{JametWTMDMP01}. Here, $ {\vec \sigma}$ is a vector
of Pauli matrices, and
\be {\vec
A}_{nm}=-\mathrm{i}\frac{\xi}{2}\sum_j\sum_{\mu,\nu}\langle\mu |
{\vec L_j}|\nu\rangle \,\psi^{}_{n}(j,\mu)\psi^{}_{m}(j,\nu)\,,
\label{Anm} \ee
where $j$ labels the atomic sites and $\mu,\nu$ the orbital bands
(\textit{s}, \textit{p}, \textit{d}). $\psi^{}_{n}(j,\mu)=\langle
j \mu |\psi^{}_{n}\rangle$ is the $n^{\rm th}$ single-particle
wavefunction, and ${\vec L}_{j}$ is the angular momentum with
respect to site $j$. The parameter $\xi$ characterizes the SO
magnitude; it is of order $80$~meV for cobalt \cite{CehovinCM02}.
The SO matrix elements satisfy ${\vec A}_{nm}= -{\vec A}_{mn}$ due
to time-reversal symmetry. We consider the case where both the
exchange energy and the single particle level spacing are large
compared with the SO matrix elements (true for small enough
nanoparticles) and treat $\Ha_{\mathrm{SO}}$ as a perturbation. In
addition, because cobalt nanoparticles apparently have a fcc
structure \cite{JametWTMDMP01}, we do not include here the
crystalline magnetic anisotropy of bulk hcp cobalt.

\begin{figure}[t]
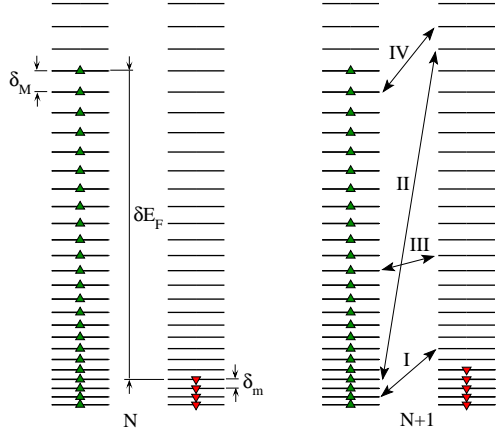

\onefigure[width=6.5cm,clip]{fig1.eps}
\caption{The single particle energy levels of $ \Ha_\mathrm{S} $ for $N$ and $N+1$ electrons.
The arrows point out different classes of transitions induced by the SO coupling, labeled I, II, III, and IV. Here $ \dm $ ($ \dM $) is the mean level spacing for the minority (majority) spins at the Fermi energy. $ \dEf $ is the single-particle energy difference between the highest occupied levels of the majority and minority spins.}
\label{scheme}
\end{figure}

We start by deriving an effective spin Hamiltonian for the nanoparticle.
In the absence of SO coupling both the occupation numbers and the total spin are good quantum numbers. In its presence,
only states with different occupation numbers are mixed (recall $\Ha_{\mathrm{SO}}$ is fully off-diagonal, ${\vec A}_{nn}=0$).
Therefore, it is natural to write an effective Hamiltonian restricted to a subspace of fixed occupation which includes virtual transitions out of
  such subspace due to SO coupling (up to second order).
This is done by a suitable canonical transformation,
$\Ha_\mathrm{eff}=
\exp[-\mathrm{\bm{i}}\hat{U}]\Ha\exp[\mathrm{\bm{i}}\hat{U}]\, .
$
Let $\{|q\rangle\}$ denote the eigenstates of $\Ha_{\mathrm{S}}$,
$\Ha_{\mathrm{S}}|q\rangle= E_q |q\rangle$. We choose
  ${\hat U}$ such that $U_{q'q}=0$ if $|q\rangle$ and $|q'\rangle$ have different occupation numbers and
\be
U_{q'q}=\mathrm{i}\frac{V_{q'q}}{(E_{q'}-
  E_q)}+ B_{q'q}
\label{U}
\ee
if they do not \cite{SlichterNMR}. Here $V_{q'q}\equiv
  (\Ha_{\mathrm{SO}})_{q'q}$ and $B_{q'q}$ is proportional to $\Ha_{\mathrm{Z}}$ and $\Ha_{\mathrm{SO}}$. As for the case of normal
metal grains \cite{GorokhovB03,CehovinCM04}, the latter leads to a correction to the g-factor which we absorb into the Zeeman term. Then, to leading order
\be
(\Ha_{\mathrm{eff}})_{q'q}= E_q\, \delta_{q'q}+(\Ha_{\mathrm{Z}}')_{q'q}\\
+\frac{1}{2}\sum_{q''}{V_{q'q''}V_{q''q}\left(\frac{1}{E_q- E_{q''}}+\frac{1}{E_{q'}- E_{q''}}\right)}
\label{Halmost}
\ee
if $|q\rangle$ and $|q'\rangle$ have the \textit{same} occupation numbers, and
$(\Ha_{\mathrm{eff}})_{q'q}=0$ if not. Using the fact
that the exchange interaction is very large ($ \dEf\!\gg\! V_{q'q} $), Eq. (\ref{Halmost}) can be further restricted
to a subspace of a given total spin $ S $. After using the Wigner-Eckart theorem
\cite{note1} and some tedious algebra, we finally obtain the following spin Hamiltonian
\be
\Ha_\mathrm{eff} = \Ha_{\mathrm{S}} + \Ha_{\mathrm{Z}}' +
   \sum_{\alpha,\beta= x,y,z}{\hat{S}_{\alpha}\, \bm{\mathcal{K}}_{\alpha\beta} \hat{S}_{\beta}}/S\,,
\label{Heff}
\ee
with an explicit expression for the anisotropy tensor $\bm{\mathcal{K}}$ given below.  Note that while $\Ha_\mathrm{eff}$ and $\Ha$ have the same eigenenergies, they do not have the same eigenvectors---in fact, $|E\rangle= \exp[\mathrm{i}\hat{U}]|E_{\mathrm{eff}}\rangle$.  Therefore, while the eigenstates of Eq. (\ref{Heff}) have well-defined spin and occupation numbers, this is \textit{not} the case for the eigenstates of the physical Hamiltonian (\ref{Hamiltonian}), as expected due to SO coupling.  The spin selection rules for electron tunneling are thus less restrictive than for a state with well-defined spin.

We classify the transitions induced by the SO coupling in four
types (see Fig.~\ref{scheme}): from doubly-occupied levels to
singly-occupied (I) or empty (II) levels, and from singly-occupied
to singly-occupied (III) or empty (IV) levels.
With this notation, the anisotropy tensor reads
\bea
\label{K}
&\bm{\mathcal{K}}_{\alpha\beta}&\smeq\sum_{m>n}A^{\alpha}_{nm}A^{\beta}_{nm}\Bigg(
\frac{-2J_S (\delta_{\rm  I}\smpl\delta_{\rm IV})}{(\ve_n\smmi\ve_m)(\ve_n\smmi\ve_m\smmi 2J_S
    S)}
\smmi\left[\frac{4J_S S}{(\ve_n\smmi\ve_m)^2\smmi4 J_S^2 S^2}\right] \frac{2\delta_{\rm III}}{(2S-1)} \\
\nonumber
&\smpl&\!\!\!\!\left[\frac{1}{\ve_n\smmi\ve_m}\frac{2}{(S+1)}\smmi\frac{1}{\ve_n\smmi\ve_m\smmi 2J_S
    S}\frac{2}{(2S+1)}
\smmi \frac{1}{\ve_n\smmi\ve_m\smpl 2J_S (S+1)}\frac{2S}{(S+1)(2S+1)}\right]\delta_{\rm II}
\Bigg)
\eea
where $\delta_X$, $X=$ I, II, III, IV, restricts the sum to the appropriate set of transitions.  It is clear from this equation that states with different occupation numbers will have different anisotropy tensors, leading to level-to-level fluctuations of the anisotropy constant. \textit{We then conclude that the addition of an electron to the nanoparticle will produce a different anisotropy tensor depending on the new occupied level}.

So far we have not specified any features of the SO coupling: Eq. (\ref{K}) is valid provided only that the SO coupling is small. A detailed treatment of the anisotropy tensor for a particular grain geometry would require the calculation of the SO matrix elements using, for instance, a non-interacting tight-binding model or a more sophisticated self-consistent Hartree-Fock approach \cite{CehovinCM02}.
However, our goal here is to calculate the generic features of the fluctuations of the anisotropy tensor. Since the shape of the grain is assumed to be irregular, we expect the SO matrix elements to fluctuate from level to level due to the fluctuations of the single-particle wavefunctions. The usual RMT approach of describing the single-particle fluctuations with the Gaussian unitary ensemble then implies that the $\{ {\vec A}_{nm} \}$ are Gaussian uncorrelated random variables because of the sum over a large number of terms in their definition Eq. (\ref{Anm}). This assumption of randomness of the SO matrix elements is consistent with the results of Ref.~\cite{CehovinCM02}.
On the other hand, we know from Refs.~\cite{DeshmukhKGBPDR01,KleffDDR01} that a uniaxial anisotropy is needed to describe the main features of the non-linear behavior in the experiment.
Therefore, we assume uniaxial anisotropy \textit{on average}:
$\langle A_x^2 \rangle \!=\! \langle A_y^2 \rangle$  and
$ \langle A_z^2 \rangle \!<\! \langle A_y^2 \rangle $.
In a more detailed calculation, this asymmetry would probably come as a difference in the energy dependence of the SO matrix elements (this is the case in Ref.~\cite{CehovinCM02}); in our model, however, it can only appear as a difference in the standard deviations.
The case of pure ``mesoscopic anisotropy''---\textit{i.e.} when all $\langle A_i^2 \rangle$ are equal---was recently analyzed in Ref.~[\citeonline{BrouwerG04}]; it was found that for $N_a \!>\! 100$ mesoscopics alone cannot explain the full magnitude of the anisotropy, thus indirectly supporting our insertion of an additional uniaxial anisotropy.
Since we do not expect $ \langle A_z^2 \rangle$ to be much smaller than $ \langle A_x^2 \rangle$, we take $ \langle A_z^2 \rangle \!=\! \langle A_x^2 \rangle/2 $; our results are only weakly sensitive to this choice (see below).

\begin{figure}[t]
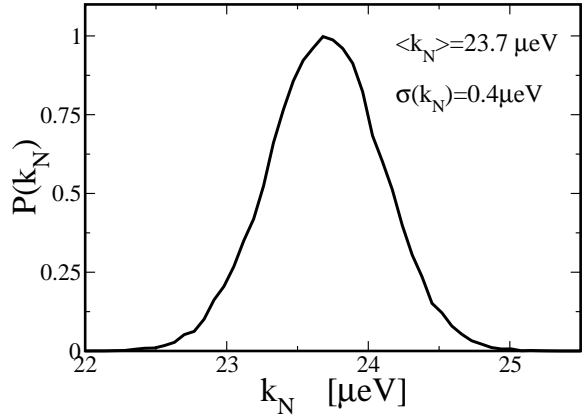

\onefigure[height=5.5cm]{fig2.eps}
\caption{Probability distribution of the magnetic anisotropy constant for
a nanoparticle size of $N_a=1200$. The distribution is Gaussian with $ \langle k_N \rangle= 23.7\, \mu$eV and $\sigma(k_N)= 0.4\, \mu$eV.}
\label{distkn}
\end{figure}

In the basis where $\bm{\mathcal{K}}$ is diagonal, the anisotropy Hamiltonian takes the form
\be
\Ha_{\mathrm{A}} =\frac{k_1 + k_2}{2S} {\hat S}^2 +
      ( k_3 - \frac{k_1 + k_2}{2}) {\hat S}_3^2/S
\ee
where $\{k_1, k_2, k_3 \}$ are the eigenvalues of $\bm{\mathcal{K}} $ and $ \hat{S}_3= \vec{S}\cdot\vec{k}_3 $. The first term is a small correction to the exchange constant and can be neglected. The second term represents the uniaxial anisotropy. We have verified that for the values used below for the SO matrix elements, $ \langle\hat{z}\!\cdot\!\vec{k}_3\rangle\!\simeq\!0.99 $, so that $ \Ha_{\mathrm{A}}\!\simeq\! -k_N \hat{S}_z^2/S $ with $ k_N= (k_1+ k_2)/2- k_3$ .  The fluctuations of $ k_N $ are calculated by numerically diagonalizing $ \bm{\mathcal{K}} $ for each realization of $ \{A_{nm}^\alpha\} $. No fluctuation is introduced for the energy levels---they are taken to be locally equally spaced, in such a way that they reproduce the (hcp) Co density of states.

The distribution of $ k_N $ for a nanoparticle containing $  N_a= 1200$ atoms is shown in Fig. \ref{distkn}, where we have used the following values for the parameters: $\dEf= 2$eV, $\dM= 3.7$meV,
$\dm= 1.15$meV,  $J_S=0.95$meV, and $\sigma(A_x) = 0.32$meV.
The latter is estimated from Eq. (\ref{Anm}) using random and uncorrelated wavefunctions,
$ \langle A_x^2\rangle = [\xi^2/27 N_a ](N_s/N_a)$---the additional factor $N_s/N_a$, with $N_s$ the number of atoms at the surface, is discussed below.
The distribution is Gaussian with a mean value $ \langle k_N \rangle= 23.7\, \mu$eV and a standard deviation $\sigma(k_N)= 0.4\, \mu$eV,  \textit{ie.} $\sigma(k_N)/\langle k_N \rangle\!\simeq\!2\%$.
These values can be estimated from Eq. (\ref{K}) by replacing the sums by integrals and considering only the diagonal terms of $ \bm{\mathcal{K}} $. We then get
\be
\langle k_N\rangle\!\simeq a \left[\langle A_x^2 \rangle-\langle A_z^2\rangle\right]\,\frac{J_S}{\dm^2}
\label{meank}
\ee
\be
\sigma(k_N)\!\simeq\! \sqrt{4 b}\left[ \left(\langle A_x^2 \rangle^2+ 2\langle A_z^2 \rangle^2\right) \ln\left(\frac{\dEf}{\dm}\right)\right]^\frac{1}{2}\frac{J_S}{\dm\,\dEf}\\
\label{rmsk} \ee where $a$ and $b$ are numerical factors that
depend on the shape of the density of states. As expected, the
fluctuations go to zero in the bulk limit, $ \sigma(k_N)/\langle
k_N\rangle\!\propto\!\dm/\dEf \!\propto\!N_a^{-1}$ up to a
logarithmic factor.  We also note that in bulk $\langle
A_x^2\rangle\!\propto\! N_a^{-1} $ so that $\langle k_N \rangle$
is independent of the system's size.  However, since we assume the
anisotropy comes from the breaking of symmetry at the surface of
the nanoparticle, we take $ \langle
A_x^2\rangle\!\propto\!N_s/N_a^2 $, so that $\langle k_N
\rangle\!\propto\!N_s/N_a$ \cite{JametWTMDMP01}. For Co, we have $
a\!\simeq\!0.8 $ and $ b\!\simeq\!1.6 $.  Evaluation of Eqs.
(\ref{meank}) and (\ref{rmsk}) gives $ \langle k_N
\rangle\!\simeq\!29\, \mu$eV and $\sigma(k_N)\!\simeq\! 0.4\,
\mu$eV, in good agreement with the numerical data.

Let us now calculate the change of the anisotropy constant upon the addition of one electron with the minority spin (right panel in Fig.~\ref{scheme}). We calculate $ k_{N+1} $ as before but take into account that both $ k_N $ and $ k_{N+1}$ contain contributions from the \textit{same} SO matrix elements. We obtain a Gaussian distribution for $\dk_N\!=\! k_{N+1}\!-\! k_{N}$, with $\langle\dk_N\rangle \!\simeq\!-0.03\,\mu$eV and $ \sigma(\dk_N)= 0.2\mu$eV.  Therefore, for a nanoparticle containing $ N_a\!\simeq\!1200 $ atoms, $\sigma(\dk_N)/\langle k_N\rangle\!\approx\! 1\% $. \textit{This is of the order needed to account for the experimental data of Ref.~\cite{DeshmukhKGBPDR01}}.

\begin{figure}[t]
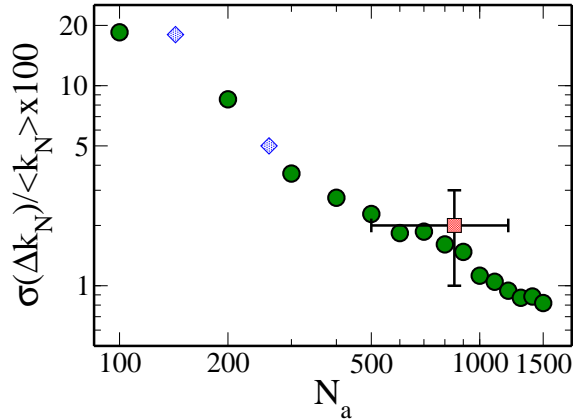

\onefigure[width=7.5cm,clip]{fig3.eps}
\caption{Relative fluctuation of the anisotropy constant upon the
  addition of one electron as a function of the numbers of atoms in
  the nanoparticles. Fluctuations are of the order of $1\%$ for
  $N_a\simeq1000$. Also shown are results of experiment (square, size of error bar corresponds to nanoparticles of $ 3$-$4$nm diameter) \cite{DeshmukhKGBPDR01,Deshmukh_Thesis} and of a self-consistent calculation (diamonds) \cite{CehovinCM02}.}
\label{deltak}
\end{figure}

Figure \ref{deltak} shows $\sigma(\dk_N)/\langle k_N\rangle$ as a function of $ N_a $. The results for the experimental data and the self-consistent calculation are indicated by the square and the diamonds, respectively.
It is important to emphasize that once the ratio $ \langle A_z^2\rangle / \langle A_x^2\rangle $ is fixed, and so is $\langle k_N\rangle$, there are no free parameters left to fit the fluctuations. Thus, the fact that the relative fluctuation of the anisotropy constant agrees reasonably well with both the experimental data and the results of a self-consistent numerical study is quite remarkable and indicates that our simple model is able to capture the essential physics.

Following the same procedure as above, we estimate
\be
\langle \dk_N\rangle \!\simeq\! c\left[\langle A_x^2 \rangle-\langle A_z^2\rangle\right] \frac{J_S}{\dm\dEf}
\label{meandk}
\ee
\be
\sigma(\dk_N)\!\simeq\!\frac{2 J_S}{\dm\dEf}\! \left(2[\langle A_x^2 \rangle^2+ 2\langle A_z^2 \rangle^2]\,\frac{1.8\dm- 1.4 J_S}{\dm- J_S}\right)^\frac{1}{2}
\label{rmsdk}
\ee
where $ c $ contains logarithmic corrections in $ \dEf/\dm $. In our case $ c\!\simeq\!-3.5 $, which gives $ \langle \dk_N\rangle\!\simeq\! -0.07\, \mu$eV and $ \sigma(\dk_N)\!\simeq\! 0.3\,\mu$eV. Note that
\be
\frac{\sigma(\dk_N)}{\langle k_N\rangle}\propto \frac{\dm}{\dEf}\propto \frac{1}{N_a}
\ee
also decays to zero in the bulk limit.  We notice that the ratio $ \sigma(\dk_N)/\langle k_N\rangle$ is \textit{independent} of the value of $ \langle A_x^2\rangle $ and only weakly dependent (up to a factor $ 2 $) on the ratio
$\langle A_z^2\rangle/ \langle A_x^2\rangle $ if the latter is $\lesssim 0.5 $. The same is true for the relative fluctuation of the ground state anisotropy, $\sigma(k_N) / \langle k_N\rangle $. Consequently, our estimate of the magnitude of the fluctuations is quite robust, being largely controlled by the ratio $ \dm/\dEf $.

It is straightforward to check from Eqs. (\ref{meandk}) and (\ref{rmsdk}) that $ \sigma(\dk_N)\!\gg\!\langle\dk_N\rangle $. This implies that $ k_{N+1} $ can be \textit{larger or smaller} than $ k_N $ with roughly equal probability.  While this is a key point for a correct account of the experimental data \cite{DeshmukhKGBPDR01}, it also has an interesting consequence if the dwell time for electrons in the nanoparticle is larger than the time it takes for the magnetization to switch. In that case, it should be possible to switch the magnetic moment of the nanoparticle by first tuning the magnetic field just below $ B_\mathrm{sw} $ for the ground state and then increasing the bias voltage until a high energy state with a smaller anisotropy constant is excited. The reverse procedure should present hysteresis. Gated nanoparticles should show similar effects as a function of the gate voltage.

Finally, let us briefly comment on the other two features of the experimental data mentioned in the introduction.  First, as noted above, $ B_{q'q}$ gives a correction to the g-factor [cf. Eq. (\ref{U})].  As for the magnetic
anisotropy, this correction will depend on the particular electronic state of the nanoparticle. We note that
even very small fluctuations will strongly modify the effective g-factor,
\be
g_{\mathrm{eff}} \simeq  \frac{|g_{N-1} S_{N-1}-g_{N} S_{N}|}{|\Delta S|} \approx g_N\,\left( 1 + \frac{S}{\Delta S}\frac{\delta g_N}{g_N} \right)
\ee
with $ \delta g_N \!=\! g_{N} \!-\! g_{N-1} $ and $ \Delta S \!=\! S_N \!-\! S_{N-1}$.  For instance, $ \delta g_N/g_N\!\simeq\!10^{-4} $, $ S\!\simeq\!1000 $ and $\Delta S \!=\! -1/2$ can lead to the small $g_{\mathrm{eff}}$ observed experimentally (this requires $ \delta g_N/\Delta S<0$, which should be verified). Second, regarding the excitation spectrum, the experimental data \cite{Deshmukh_Thesis} suggest that the low energy spacing is roughly independent of the grain's size \cite{note2}. The only quantity with such scaling is the anisotropy constant. Then, resonances should be related to the excitation of collectives modes (magnetization) \cite{CehovinCM03} along an axis with a much larger anisotropy (hard axis)---this is crucial since the smallness of $ B_\mathrm{sw} $ implies that the anisotropy along the easy axis is also small. We believe such resonances are described by a more general form of the anisotropy, such as $ \Ha_\mathrm{A}= -k S_z^2+ k' S_x^2 $. Work in this direction is in progress.

\acknowledgments
We thank M. Deshmukh and D. Ralph for many valuable conversations and appreciate helpful discussions with C. Canali and A. MacDonald.  GU acknowledges support from CONICET (Argentina) and Fundaci\'on Antorchas (Grant 14169/21). This work was supported in part by the NSF (DMR-0103003) and ANPCyT Grant No 13476.

%%%%%%%%%%%%%%%%%%%%%%%%%%%%%%%%%%%%%%%%%%%%%%%%%%%%%%%%%%%%


\begin{thebibliography}{16}

\bibitem{GueronDMR99}
\Name{Gu{\'e}ron S., Deshmukh M.~M., Myers E.~B. \and Ralph D.~C.}
\REVIEW{Phys. Rev. Lett.}{83}{1999}{4148}.

\bibitem{DeshmukhKGBPDR01}
\Name{Deshmukh M.~M., Kleff S., Gu\'eron S., Bonet E., Pasupathy A. N., von Delft J. \and Ralph D. C.}
\REVIEW{Phys. Rev. Lett.}{87}{2001}{226801}

\bibitem{vanDelftR01}
\Name{von Delft J. \and Ralph D.~C.}
\REVIEW{Phys. Rep.}{345}{2001}{62}

\bibitem{KleffDDR01}
\Name{Kleff S., von Delft J., Deshmukh M.~M. \and Ralph D.~C.}
\REVIEW{Phys. Rev. B}{64}{2001}{220401}

\bibitem{CanaliM00}
\Name{Canali C.~M. \and MacDonald A.~H.}
\REVIEW{Phys. Rev. Lett.}{85}{2000}{5623}

\bibitem{KleffD02}
\Name{Kleff S. \and von Delft J.}
\REVIEW{Phys. Rev. B}{65}{2002}{214421}

\bibitem{CehovinCM03}
\Name{Cehovin A., Canali C.~M. \and MacDonald A.~H.}
\REVIEW{Phys. Rev. B}{68}{2003}{014423}

\bibitem{CehovinCM02}
\Name{Cehovin A., Canali C.~M. \and MacDonald A.~H.}
\REVIEW{Phys. Rev. B}{66}{2002}{094430}

\bibitem{BrouwerG04}
\Name{Brouwer P. W. \and Gorokhov D. A.}
\REVIEW{Phys. Rev. Lett.}{95}{2005}{017202}

\bibitem{JametWTMDMP01}
\Name{Jamet M., Wernsdorfer W., Thirion C., Dupuis V., Melinon P., Perez A. \and Mailly D.}
\REVIEW{Phys. Rev. Lett.}{86}{2001}{4676}

\bibitem{SlichterNMR}
\Name{Slichter C.~P.}
\Book{Principles of Magnetic Resonance}
\Publ{Springer-Verlag, New York}
\Year{1990}

\bibitem{GorokhovB03}
\Name{Gorokhov D.~A. \and Brouwer P.~W.}
\REVIEW{Phys. Rev. Lett.}{91}{2003}{186602}

\bibitem{CehovinCM04}
\Name{Cehovin A., Canali C.~M. \and MacDonald A.~H.}
\REVIEW{Phys. Rev. B}{69}{2004}{045411}

\bibitem{note1}
To evaluate the matrix element $V_{q'q}$, we define the operators
  $\mathcal{O}^1_1(n,m)=-c^{\dagger}_{n\up}c^{}_{m\dn}$,
  $\mathcal{O}^1_{-1}(n,m)= c^{\dagger}_{n\dn}c^{}_{m\up}$ and
  $\mathcal{O}^1_{0}(n,m)= (c^{\dagger}_{n\up}c^{}_{m\up}-
  c^{\dagger}_{n\dn}c^{}_{m\dn})/\sqrt{2}$ and use the Wigner-Eckart theorem:
  $\langle q'|\mathcal{O}^1_k(n,m)|q\rangle= (-1)^{S'-S'_z}$ $
  C_{3j}(\{S',S'_z\},\{1,k\},\{S,S_z\})\, \langle
  q'||\mathcal{O}^1(n,m)||q\rangle$ where $C_{3j}$ is the $3$j-symbol. A
  general expression for the reduced matrix element $\langle
  q'||\mathcal{O}^1(n,m)||q\rangle$ can be found in Ref. \cite{GouyetSS75}.
  We restrict our calculation to states with maximum spin.

\bibitem{Deshmukh_Thesis}
\Name{Deshmukh M.~M.}
Ph.D. thesis, Cornell University (2002).

\bibitem{note2}
 Our results concern particle-hole excitations, which cannot account for all the resonances observed in Refs.
  \cite{GueronDMR99,DeshmukhKGBPDR01}.

\bibitem{GouyetSS75}
\Name{Gouyet J.~F., Schranner R. \and Seligman T.~H.}
\REVIEW{J. Phys. A: Math. and Gen.}{8}{1975}{285}


\end{thebibliography}
\end{document}